\documentclass[12pt, a4, draftclsnofoot, onecolumn]{IEEEtran}

\usepackage[T1]{fontenc}

\usepackage[none]{hyphenat}
\usepackage[utf8]{inputenc}
\usepackage[english]{babel}

\usepackage{verbatim}
\usepackage{graphicx}
\usepackage{float}
\usepackage{booktabs}
\usepackage{mathtools}
\usepackage{hyperref}
\usepackage{lipsum}
\usepackage{stfloats}
\usepackage{amsmath,amsthm}
\usepackage{amssymb}
\usepackage{mathtools}
\usepackage{epstopdf}
\usepackage{booktabs,dcolumn,caption}
\usepackage{multirow}
\usepackage{caption}
\usepackage{subcaption}
\usepackage{ragged2e}
\usepackage{cite}
\usepackage{censor}
\usepackage{tabularx}
\usepackage{relsize}

\usepackage{booktabs} 
\usepackage{siunitx}  
\usepackage[margin=1in]{geometry} 

\usepackage{array}
\newcolumntype{C}[1]{>{\centering\arraybackslash}p{#1}}
\newcolumntype{L}[1]{>{\RaggedRight\arraybackslash}p{#1}}
\usepackage{wasysym}

\newcommand*\diff{\mathop{}\!\mathrm{d}}

\usepackage{mathrsfs}
\usepackage{xcolor}
\usepackage{colortbl}
\usepackage{color,soul}

\usepackage{caption,hypcap}

\usepackage{cleveref}

\usepackage{suffix}
\usepackage{mathtools, cuted}
\usepackage{lipsum, color}

\usepackage{enumitem}

\DeclarePairedDelimiterX\MeijerM[3]{\lparen}{\rparen}%
{#3\delimsize\vert\,\begin{matrix}#1 \\ #2\end{matrix}}

\newcommand\MeijerG[8][]{%
	G^{\,#2,#3}_{#4,#5}\MeijerM[#1]{#6}{#7}{#8}}

\WithSuffix\newcommand\MeijerG*[7]{%
	G^{\,#1,#2}_{#3,#4}\MeijerM*{#5}{#6}{#7}}

\usepackage{array,collcell}
\newcommand\AddLabel[1]{%
  \refstepcounter{equation}
  (\theequation)
  \label{#1}
}
\newcolumntype{M}{>{\hfil$\displaystyle}X<{$\hfil}} 
\newcolumntype{L}{>{\collectcell\AddLabel}r<{\endcollectcell}}

\hypersetup{linkcolor=black,citecolor=black}

\usepackage{mathtools, cuted}

\begin{document}

\title{Error performance analysis of different modulations over the TWDP fading channels}
\title{Error performance analysis of different modulations over TWDP fading channel}

\author{Almir~Maric
,~\IEEEmembership{Member,~IEEE,}
        and~Pamela~Njemcevic
        ,~\IEEEmembership{Member,~IEEE}
\thanks{A. Maric and P. Njemcevic are with the Department
of Telecommunications, Faculty of Electrical Engineering, University of Sarajevo, Sarajevo, Bosnia and Herzegovina.}
\thanks{e-mail: almir.maric@etf.unsa.ba;}
}

%
%


\maketitle

\begin{abstract}
Two-wave with diffuse power (TWDP) is one of the most promising distribution for description of a small-scale fading in the emerging mmWave band. However, traditional error performance analysis in these channels faces two  fundamental issues. It is mostly based on conventional TWDP  parameterization which is not  in accordance with the model’s  underlying physical mechanisms and which hinders accurate observation of the impact of a model parameters on a system’s performance metrics. In addition, the existing average bit/symbol error probability (ABEP/ASEP) expressions for most modulations and diversity schemes are available as approximations, which are accurate only for specific combinations of TWDP parameters.
Accordingly, in this paper, the exact ASEP expressions are derived for M-ary rectangular quadrature amplitude modulation (RQAM) with coherent detection and for M-ary DPSK modulation, and are given in terms of physically justified parameters. Besides, in order to relax computational complexity of proposed exact ASEPs in high signal-to-noise ratio (SNR) region, their asymptotic counterparts are derived as the simple closed-form expressions, matching the exact ones for SNR>30dB.
Results are verified by Monte-Carlo simulation.
\end{abstract}
\begin{IEEEkeywords}
TWDP fading channel, ASEP, RQAM, DPSK.
\end{IEEEkeywords}

\IEEEpeerreviewmaketitle

\section{Introduction}
\label{sec:I}
TWDP is a small-scale fading model, introduced by Durgin~\textit{et al.} for modeling Rician, Rayleigh and hyper-Rayleigh fading conditions~\cite{Dur02}. It assumes that received signal's complex envelope is composed of two strong specular components and many weak diffuse components. It is empirically shown that such a signal occurs in 5G networks, where mmWave band and directional antennas are employed~\cite{Zoc19, Zoc19-1, Mav15}, as well as in wireless sensor networks deployed in cavity environments such as airplanes, buses, etc.~\cite{Fro07, Fro08}.

Although applicable for modeling of received signals in different emerging technologies, TWDP model from its inception has been facing two major issues. First, initial TWDP probability density function (PDF) expressions are given in integral or approximate forms, which hinders the exact wireless system performance analysis in channels with TWDP fading~\cite{Rao15}. Besides, conventional TWDP parameterization is shown to be inconsistent with model's underlying physical mechanisms, which complicates observation of the impact of the ration between different signal components
on a system’s performance metrics~\cite{rad}.

Namely, the original TWDP envelope PDF~\cite[eq. (32)]{Dur02} is given in integral form, which couldn't be used for further mathematical manipulations. So, as a solution, Durgin~\textit{et al.} proposed the closed-form approximation of TWDP envelope's PDF~\cite[eq. (17)]{Dur02}, used in many papers to evaluate error performance of wireless communication systems with the single- and multi-antenna reception.
Accordingly, average symbol/bit error probabilities are derived for different modulations and diversity schemes, either by averaging conditional BEP/SEP in channel with additive white Gaussian noise (AWGN) over the approximate TWDP PDF~\cite[eq. (17)]{Dur02} (PDF approach), or based on 
the approximate moment generation function (MGF)~\cite[eq. (11)]{Oh07} also derived from~\cite[eq. (17)]{Dur02} (MGF approach). Comprehensive overview of these results, classified according to derivation approach and according to treated modulation and diversity techniques, is given within the first two columns of Table~\ref{Tab:01}. 
\begin{table}[]
\centering
\caption{An overview of available average error probability expressions for different modulation and diversity techniques in TWDP channel}
\begin{tabular}{ll||c|c|c}
\toprule
\multicolumn{2}{c}{\multirow{2}{*}{\begin{tabular}[c]{@{}c@{}}Modulation\\ (coherent detection)\end{tabular}}} & \multicolumn{2}{c}{\textit{Approx. expressions}}   & \multicolumn{1}{c}{\multirow{2}{*}{\begin{tabular}[c]{@{}c@{}}\textit{Exact}\\ \textit{expressions}\end{tabular}}} \\
\multicolumn{2}{c}{}                            & \multicolumn{1}{c}{\begin{tabular}[c]{@{}c@{}}PDF\\ approach\end{tabular}} & \multicolumn{1}{c}{\begin{tabular}[c]{@{}c@{}}MGF\\ approach\end{tabular}} & \multicolumn{1}{c}{}                                                                             \\
\midrule
BPSK       & no diversity   & {\cite{Oh05}}     & {}                & {\cite{rad}}\\
           & L-branch SC    & {\cite{Sub13}}    & {}                & {}\\
           & L-branch EGC   & {\cite{Das17}}    & {}                & {}\\
           & L-branch MRC   & {\cite{Sub14}}    & {\cite{Oh07}}     & {}\\
\midrule
M-ary      & no diversity   & {\cite{Sin15}}    & {}                & {\cite{rad}}\\
PSK        & 2-branch SC    & {}                & {\cite{Tan11}}    & {}\\
           & L-branch MRC   & {\cite{Sub14}}    & {}                & {}\\
\midrule
M-ary      & no diversity   & {}                & {\cite{Dix13}}    & {}\\
SQAM       & L-branch MRC   & {\cite{Sub14}}    & {}                & {}\\
\midrule
M-ary      & no diversity   & {\cite{Sur08}}    & {\cite{Dix13}}    & {}\\
RQAM       & 2-branch SC    & {}                & {\cite{Tan11}}    & {}\\
           & L-branch SC    & {\cite{Das17-1}}  & {}                & {}\\
           & L-branch MRC   & {}                & {\cite{Lu12}}     & {}\\
\bottomrule
\toprule
\multicolumn{2}{c}{\multirow{2}{*}{\begin{tabular}[c]{@{}c@{}}Modulation\\ (noncoherent detection)\end{tabular}}} & \multicolumn{2}{c}{\textit{Approx. expressions}}   & \multicolumn{1}{c}{\multirow{2}{*}{\begin{tabular}[c]{@{}c@{}}\textit{Exact}\\ \textit{expressions}\end{tabular}}} \\
\multicolumn{2}{c}{}                            & \multicolumn{1}{c}{\begin{tabular}[c]{@{}c@{}}PDF\\ approach\end{tabular}} & \multicolumn{1}{c}{\begin{tabular}[c]{@{}c@{}}MGF\\ approach\end{tabular}} & \multicolumn{1}{c}{}                                                                             \\
\midrule
BFSK        & no diversity  & {}                & {}                & {\cite{Rao15}}\\
            & 2-branch SSC  & {\cite{Hag11}}    & {\cite{Lee07}}    & {}\\
            & L-branch SC   & {\cite{Sub13}}    & {}                & {}\\
            & L-branch EGC  & {}                & {\cite{Lee07-1}}  & {}\\
            & L-branch MRC  & {\cite{Sub14}}    & {}                & {}\\ 
\midrule
M-ary       & no diversity & {}                & {}                & {\cite{Rao15}}\\
FSK         & 2-branch SSC  & {\cite{Hag12}}    & {}                & {}\\
            & L-branch EGC          & {}                & {}                & {\cite{Mar18}}\\           
\midrule
DBPSK       & no diversity & {}                & {}                & {\cite{Rao14}}\\
            & L-branch SC   & {\cite{Sub13}}    & {}                & {}\\
            & L-branch EGC  & {}                & {\cite{Lee07-1}}  & {}\\
\midrule
M-ary       & L-branch MRC  & {\cite{Sub14}}    & {}                & {}\\
DPSK        & {}            & {}                & {}                & {}\\
\bottomrule
\end{tabular}
\label{Tab:01}
\end{table}

Overviewed papers have obviously provided the first analytical insight in ABEP/ASEP for channels with TWDP fading in a number of scenarios~\cite{Rao15}. However, obtained results are approximations and their accuracy is known to degrade as the power of two specular components become stronger in respect to the power of diffuse components, while simultaneously two specular components become similar in strength~\cite{Dur02}. Besides, they are all based on conventional parameterization, which disables clear observation of the effect of  different ratios between specular components on  ABEP/ASEP values~\cite{rad}. 

Step forward in error performance analysis over the TWDP fading channels had been made in~\cite{Rao15}, by derivation of a closed-form TWDP MGF expression~\cite[eq. (25)]{Rao15}. 
However, although exact, it is shown that proposed MGF expression is not appropriate for mathematical manipulations. Accordingly, it is used only for derivation of the exact error probability of binary DPSK~\cite{Rao14, Rao15} and noncoherent M-ary frequency shift keying (FSK) modulations without diversity~\cite{Rao15} and with the equal gain combining (EGC) reception~\cite{Mar18}, also in terms of conventional TWDP parameters. 

However, improved TWDP parameterization that is in accordance with model’s underlying physical mechanisms is recently proposed in~\cite{rad}. Besides, the alternative form of the exact  MGF~\cite[eq. (22)]{rad} is also derived based on the exact PDF~\cite[eq. (6)]{Kos78}, and expressed in terms of improved parameters. It is shown that proposed form of MGF is suitable for mathematical manipulations and demonstratively, it is used to derive the exact ASEP for M-ary PSK system with the single-antenna coherent detection~\cite{rad}. 

So, despite the sporadic breakthroughs in TWDP error performance analysis, the exact error performance characterization which allows accurate observation of the impact of model parameters on a system’s performance metrics, for most modulation and diversity schemes in TWDP channels, remains an open issue~\cite{Rao15}. 

Accordingly, in this paper the exact ASEP expression for a generic, coherently detected  M-ary RQAM system is derived and used for derivation of the exact ASEP for M-ary SQAM and M-ary amplitude shift keying (ASK), as well as for binary and quadrature PSK modulations, all in terms of physically justified TWDP parameters. The exact ASEP expression for M-ary DPSK modulated signal is also derived. The error performance of all considered modulations are then compared, revealing superiority of SQAM modulation and explaining reason for its extensive applicability in 5G and other networks in which multipath fading can be best described using TWDP model. 
In addition, for all considered modulations, high SNR asymptotic ASEPs are also derived as a simple closed-form expressions, enabling us to relax computational complexity of their exact counterparts in high SNR region.

The remainder of the paper is structured as follows: In Section \ref{sec:II}, TWDP channel model is presented. In Section \ref{sec:III}, the exact and asymptotic ASEP expressions for M-ary modulations with coherent and differential detection are derived. Results are verified in Section \ref{sec:IV} in terms of Monte-Carlo simulation. Conclusion is provided in Section \ref{sec:V}. 

\section{TWDP fading model}
\label{sec:II}
TWDP fading model, introduced in~\cite{Dur02}, is useful for modeling many practical wireless communication systems. It assumes that in the frequency nonselective slow fading channel, the complex envelope $r(t)$ of the received signal is composed of two strong specular components $v_1(t)$ and $v_2(t)$ and many low-power diffuse components treated as a random process $n(t)$. Thus, $r(t)$ can be expresses as:
\begin{equation}
\begin{split}
    r(t) & =  v_1(t) + v_2(t) + n(t) \\ & = V_1\exp{\left(j\Phi_1\right)} + V_2\exp{\left(j\Phi_2\right)} + n(t)
    \end{split}
    \label{eq01}
\end{equation} 
In (\ref{eq01}) specular components are assumed to have constant magnitudes $V_1$ and $V_2$ and uniformly distributed phases $\Phi_1$ and $\Phi_2$ in $[0, 2\pi)$, while diffuse components are treated as a complex zero-mean Gaussian random process $n(t)$ with the average power $2\sigma^2$. Consequently, the average power of $r(t)$ equals to $\Omega = V_1^2 + V_2^2 + 2\sigma^2$. Besides the average power $\Omega$, two additional parameters $K$ and $\Delta$, defined in~\cite{Dur02} as $K = (V_1^2+V_2^2)/(2\sigma^2)$ and $\Delta = V_1V_2/(V_1^2+V_2^2)$ respectively, are traditionally used for TWDP model description. However, it is shown in~\cite{rad} that conventional $\Delta$-based parameterization is not in accordance with model’s underlying physical mechanisms.
Consequently, its usage hinders
accurate observation of the impact of the ratio between specular components on a system’s performance metrics~\cite{rad} and causes anomalies related to the estimation of these parameters~\cite{rad1}. To overcome the issues, improved TWDP parameterization, based on physically justified parameters $K$ and $\Gamma$ is introduced in~\cite{rad}: 
\begin{equation}
\label{eq02}
   K = \frac{V_1^2+V_2^2}{2 \sigma^2},~~~~~~\Gamma = \frac{ V_2}{V_1} 
\end{equation}
where $K$ reflects the ratio between the powers of specular and diffuse components and $\Gamma$ reflects the ratio between magnitudes of two specular components. Since it is shown that no anomalies related to the observation of the impact of a model parameters on ASEP~\cite{rad} and to parameter estimation~\cite{rad1} can be observed when using improved parameters, parameterization based on $K$ and $\Gamma$ is used for description of TWDP fading in the remainder of this paper. 

So, considering adopted parameterization and the overall model's assumptions, TWDP fading is properly described by its envelope PDF recently obtained in~\cite{rad} based on~\cite{Kos78}, as the exact infinite-series expression:  
\begin{equation}
	\label{eq02'}
	\begin{split}
	    f_R&(r)=\frac{r}{\sigma^2}\exp{\left(-\frac{r^2 }{2\sigma^2}-K\right)} \sum_{m=0}^{\infty} \varepsilon_m (-1)^m  I_m\left(2r\sqrt{\frac{K}{2\sigma^2}\frac{1}{1+\Gamma^2}}\right) \\ & \times I_m\left(2r \sqrt{\frac{K}{2\sigma^2}\frac{\Gamma^2}{1+\Gamma^2}}\right)
        I_m\left(2K\frac{\Gamma}{1+\Gamma^2}\right)
	\end{split}
\end{equation}
where $\varepsilon_0 = 1$, $\varepsilon_m = 2$ for $m \geq 1$ and $I_{\nu}(\cdot)$ is a modified $\nu$-th order Bessel function of the first kind.
That expression is then used to derive the alternative form of the exact MGF of the SNR, given as~\cite{rad}: 
\begin{equation}
	\label{eq03}
	\begin{split}
	    \mathcal{M}_\gamma&(s) = 		\frac{1+K}{1+K-s\gamma_0}		\sum_{m=0}^{\infty} \frac{1}{m!} \left(\frac{K}{1+\Gamma^2}\right)^m \times \left(\frac{\gamma_0s}{1+K-s\gamma_0}\right)^m   {}_2F_1\left(-m,-m;1;\Gamma^2\right)
	\end{split}
\end{equation}
where \mbox{$\gamma_0=2\sigma^2 (1+K)\frac{T_s}{N_0}$} is the average SNR, $T_s$ denotes  symbol time, $N_0/2$ is the power spectral density of the white Gaussian noise and $_2 F_1(\cdot,\cdot;\cdot; \cdot)$ is the Gaussian  hypergeometric function.

It is shown in~\cite{rad} that the alternative form of the MGF given by (\ref{eq03}) can be reduced to the form of the MGF proposed in~\cite{Rao15}. However, according to~\cite{rad}, this alternative form is much more suitable for mathematical manipulations. Since it is also given in terms of physically justified parameters, (\ref{eq03}) is used as a basis for ASEP derivations in following section. 

\section{Error probability performance analysis over TWDP channels}
\label{sec:III}
In wireless communication systems, accurate and tractable analysis of different system performance metrics is of profound significance for network planning and reliable transceiver design. Among these metrics, symbol error probability is one of the most important, since it enables us to quantify performance of systems with different modulations, when they are subjected to various TWDP fading conditions~\cite{Sur08}. However, according to the overview provided in Section~\ref{sec:I}, it is obvious that in considered channel, ABEP/ASEPs for most modulations are available only as approximate expressions given in term of nonphysical parameters, with the accuracy limited 
by the accuracy of employed approximate PDF expression. 

\subsection{Exact average symbol error probability expressions}

According to the overview provided in Section~\ref{sec:I}, the 
exact ASEP expressions in TWDP fading channels are available only for noncoherent M-ary FSK and differential BPSK modulations, as well as for M-ary PSK with a single-antenna coherent detection, while  for all other modulations and diversity schemes, the exact  expressions are missing. In additions, only the exact M-ary PSK ASEP is given in terms of parameters $K$ and $\Gamma$. Accordingly, the following section closes the gap, by providing the exact ASEP expressions for remaining most popular M-ary modulation schemes in systems with the single-antenna reception,  given in terms of physically justified $K$ and $\Gamma$. 


\subsubsection{M-ary RQAM}
The general order rectangular QAM is equivalent to two independent pulse amplitude modulation (PAM) signals ($M_I$-PAM in-phase and $M_Q$-PAM quadrature signals)~\cite{Sur08}. 
Accordingly, the integral M-ary RQAM ASEP with the order of modulation $M = M_I \times M_Q$, can be expressed as~\cite{Lu11}:
\begin{equation}
\begin{split}
 \label{eq04}
    P_s&{^{RQAM}(\gamma_0)} =   a_I  \int\displaylimits_{0}^{\frac{\pi}{2}} \mathcal{M}_\gamma\left(\frac{A_I^2}{2\sin^2{\theta}}\right) \diff{\theta}  +  a_Q  \int\displaylimits_{0}^{\frac{\pi}{2}} \mathcal{M}_\gamma\left(\frac{A_Q^2}{2\sin^2{\theta}}\right) \diff{\theta} \\
    & - \frac{\pi}{2} a_I a_Q\int\displaylimits_{0}^{\frac{\pi}{2}-\tan^{-1}{\frac{A_Q}{A_I}}} \mathcal{M}_\gamma\left(\frac{A_I^2}{2\sin^2{\theta}}\right) \diff{\theta}- \frac{\pi}{2} a_I a_Q\int\displaylimits_{0}^{\tan^{-1}{\frac{A_Q}{A_I}}} \mathcal{M}_\gamma\left(\frac{A_Q^2}{2\sin^2{\theta}}\right) \diff{\theta}
    \end{split}
\end{equation}
where 
\begin{equation}
\nonumber
\begin{split}
    A_I &= \sqrt{\frac{6}{(M_I^2-1) + \beta^2 (M_Q^2-1)}},~a_I = \frac{2}{\pi}\frac{M_I-1}{M_I}\\
    A_Q &= \beta A_I,~~~~~~~~~~~~~~~~~~~~~~~~~~~~a_Q = \frac{2}{\pi}\frac{M_Q-1}{M_Q}
    \end{split}
\end{equation}
while $\beta$ represents the ratio between quadrature and in-phase decision distances, i.e. $\beta = d_Q/d_I$. So, after inserting (\ref{eq03}) in (\ref{eq04}), the exact ASEP of M-ary RQAM can be expressed as:
\begin{equation}
\begin{split}
 \label{eq04b}
    P_s&^{RQAM}(\gamma_0) =  \sum_{m=0}^{\infty} \left(\frac{K}{1+\Gamma^2}\right)^m {}_2F_1\left(-m,-m;1;\Gamma^2\right) \times \frac{1}{m!} \left[a_I \mathcal{I}\left(\frac{\gamma_0}{1+K},A_I,\frac{\pi}{2}\right)\right.\\
    &\left.+a_Q \mathcal{I}\left(\frac{\gamma_0}{1+K},A_Q,\frac{\pi}{2}\right)-\frac{\pi}{2}a_I a_Q \mathcal{I}\left(\frac{\gamma_0}{1+K},A_I,\frac{\pi}{2}-\tan^{-1}{\frac{A_Q}{A_I}}\right)\right.\\
    &\left.-\frac{\pi}{2}a_I a_Q \mathcal{I}\left(\frac{\gamma_0}{1+K},A_Q,\tan^{-1}{\frac{A_Q}{A_I}}\right)\right]
		%
    \end{split}
\end{equation}
where $\mathcal{I}(\cdot,\cdot,\cdot)$ is defined as:
\begin{equation}
\begin{split}
 \label{eq100}
    \mathcal{I}(a,A,\vartheta)=\int\displaylimits_{0}^{\vartheta} \frac{1}{1- a s} \left(\frac{a s}{1-a s}\right)^m \diff{\theta}\Bigg\vert_{s=\frac{A^2}{2\sin^2{\theta}}}
		%
    \end{split}
\end{equation}
Integral in (\ref{eq100}) can be solved in terms of Appell's hypergeometric function $F_1(\cdot;\cdot,\cdot; \cdot;\cdot,\cdot)$ as:
\begin{equation}
\begin{split}
    \mathcal{I}(a,A,&\vartheta) = (-1)^{m+1}\frac{2}{3}  \frac{\sin ^3{\vartheta}}{a A^2} \times F_1\left(\frac{3}{2};m+1,\frac{1}{2};\frac{5}{2};\frac{2 \sin ^2{\vartheta}}{a A^2},\sin ^2{\vartheta}\right)
    \end{split}
\end{equation}
with special cases:
\begin{equation}
\nonumber
\begin{split}
    \mathcal{I}(a,A,\frac{\pi}{2}) &= (-1)^{m+1} \frac{\pi}{2 a A^2} {}_2 F_1\left(\frac{3}{2};m+1;2;\frac{2}{a A^2}\right)
    \end{split}
\end{equation}
\begin{equation}
\nonumber
\begin{split}
    \mathcal{I}(a,A,&\tan^{-1}\phi) = (-1)^{m+1} \frac{2}{3 a A^2} \left(\frac{\phi^2}{1+\phi^2}\right)^{\frac{3}{2}} \times F_1\left(\frac{3}{2};m+1,\frac{1}{2};\frac{5}{2};\frac{2}{a A^2} \frac{\phi^2}{1+\phi^2},\frac{\phi^2}{1+\phi^2}\right)
    \end{split}
\end{equation}
\begin{equation}
\nonumber
\begin{split}
    \mathcal{I}(a,A,&\frac{\pi}{2}-\tan^{-1}\phi) = (-1)^{m+1} \frac{2}{3 a A^2} \left(\frac{1}{1+\phi^2}\right)^{\frac{3}{2}} \times F_1\left(\frac{3}{2};m+1,\frac{1}{2};\frac{5}{2};\frac{2}{a A^2} \frac{1}{1+\phi^2},\frac{1}{1+\phi^2}\right)
    \end{split}
\end{equation}
Considering the above given integral solution, the exact ASEP of a single-branch M-ary RQAM receiver is finally obtained as the exact expression, given by (\ref{eq05}) in Table~\ref{Tab:02}.

\begin{table}[b]
\centering
\caption{M-ary RQAM: Special cases}
\begin{tabular}{l|ccccc}
\toprule
\multicolumn{1}{c}{\multirow{2}{*}{Modulation}} & \multicolumn{5}{c}{RQAM parameters}                       \\
\multicolumn{1}{c}{}                            & $M_I$ & $M_Q$ & $A_I$ & $A_Q$ & $\beta$ \\
\midrule
M-ary SQAM
& $\sqrt{M}$ & $\sqrt{M}$ & $\sqrt{\frac{3}{M-1}}$ & $\sqrt{\frac{3}{M-1}}$ & 1 \\
M-ary ASK                & $M$ & $1$ & $\sqrt{\frac{6}{M^2-1}}$ & $\sqrt{\frac{6}{M^2-1}}$ & $1$ \\
QPSK                & $2$ & $2$ & $1$ & $1$ & $1$ \\
BPSK                & $2$ & $1$ & $\sqrt{2}$ & $\sqrt{2}$ & $1$ \\
\bottomrule
\end{tabular}
\label{table:RQAM}
\end{table}



\subsubsection{M-ary SQAM, M-ary ASK, QPSK and BPSK} Rectangular QAM represents a generic modulation scheme since it includes SQAM, ASK, quadrature PSK and binary PSK schemes as its special cases. Accordingly, ASEP expressions for listed modulations can be obtained from (\ref{eq05}) by employing specific values of RQAM parameters for each considered modulation (see Table~\ref{table:RQAM}). 
Accordingly, ASEP for M-ary SQAM, M-ary ASK, QPSK and BPSK are listed in Table~\ref{Tab:02} and labeled by (\ref{eq06}) - (\ref{eq09}), respectively.
It is also note worthy that the derived ASEPs for quadrature and binary PSK are exactly the same as the corresponding expressions obtained from M-ary PSK ASEP~\cite[eq. (28)]{rad} for $M = 4$ and $M = 2$ and accordingly, results based on these expressions are not going to be further elaborated.  

It is also important to emphasize that among given results, those related to SQAM are of upmost importance, since SQAM presents a preferred
modulation scheme in the emerging 5G networks due to its high bandwidth efficiency and implementation simplicity~\cite{Dix13}.

\subsubsection{M-ary DPSK}
In order to derive the exact M-ary DPSK ASEP, lets start from its integral counterpart, given in~\cite{Rao15} as:
\begin{equation}
\label{eq10}
    P_s^{DPSK} = \frac{1}{\pi} \int\displaylimits_{0}^{\left(1-\frac{1}{M}\right)\pi}
    \mathcal{M}_\gamma\left(\frac{-\sin^2{\frac{\pi}{M}}}{1 + \cos{\frac{\pi}{M}}\cos{\theta}}\right) \diff{\theta}
\end{equation}
which can be, after inserting (\ref{eq03}) in (\ref{eq10}) and some manipulations, expressed as:
\begin{equation}
\begin{split}
    \label{eq10b}
    P_s^{DPSK}&(\gamma_0) =  \sum_{m=0}^{\infty} \frac{1}{m!} \left(\frac{K}{1+\Gamma^2}\right)^m \times {}_2F_1\left(-m,-m;1;\Gamma^2\right)\left(\mathcal{I}_m + \mathcal{I}_{m+1}\right)
\end{split}
\end{equation}
where $\mathcal{I}_{(\cdot)}$ is defined as:
\begin{equation}
    \begin{aligned}
        \mathcal{I}_m=-\frac{1}{\pi} \int\displaylimits_{0}^{\left(1-\frac{1}{M}\right)\pi} \left(1 + 
         \frac{1+ \cos{\frac{\pi }{M}} \cos{\theta}}{a(1-\cos ^2{\frac{\pi }{M}})}\right)^{-m} \diff{\theta}
    \end{aligned}
\end{equation}

\noindent By using binomial formula and after some manipulations, $\mathcal{I}_m$ can be solved as:
\begin{equation}
    \label{10c}
    \begin{split}
    &\mathcal{I}_m = \sum _{p=0}^{\infty } \binom{-m}{p} \left(\frac{b-A b}{A}\right)^p \left(\frac{A}{a-b^2}\right)^{m+p} \\
    & \times \left(\frac{\Gamma \left[\frac{p+1}{2}\right]}{2 \sqrt{\pi } \Gamma \left[\frac{p+2}{2}\right]}-\frac{\left(-b\right)^{p+1}}{\pi  (p+1)}\right.\\
    &\left.\times{}_2F_1\left(\frac{1}{2},\frac{p+1}{2};\frac{p+3}{2};b^2\right)\right)
    \end{split}
\end{equation}
where $A = \frac{\gamma_0(1-b^2)}{1+K+\gamma_0(1-b^2)}$, $b = \cos{\frac{\pi}{M}}$, $a = \frac{\gamma_0}{1+K}$ and $\Gamma\left[\cdot \right]$ is the Gamma function. Accordingly, the exact M-ary DPSK ASEP expression can be determined as (\ref{eq10e}), given in Table~\ref{Tab:02}.

It is important to emphasize that by using~\cite[eq. (23)]{rad} and by employing the relation between parameters $\Gamma$ and $\Delta$ (i.e. $\frac{2\Gamma}{\Gamma^2+1} = \Delta$), expression (\ref{eq10e}) for $M=2$ can be reduced to 
the exactly the same expressions as the one derived in~\cite{Rao15} by directly solving integral (\ref{eq10}) for the same value of $M$.

\begin{table*}
\caption{Exact ASEP expressions} 
\centering
\begin{tabular}[t]{l|lL}
\toprule
\multicolumn{1}{c}{Modulation} & \multicolumn{2}{l}{ASEP} \\
\midrule
\begin{tabular}[c]{@{}l@{}}M-ary\\ RQAM\end{tabular} & 
$\!\begin{aligned}[t]
        P_s^{RQAM}(\gamma_0) &= \frac{2}{\pi} \frac{1+K}{\gamma_0} \sum_{m=0}^{\infty} \frac{(-1)^{m+1}}{m!}        \left(\frac{K}{1+\Gamma^2}\right)^m {}_2F_1\left(-m,-m;1;\Gamma^2\right)   \\
        & \times \left\{\frac{2(M_I-1)(M_Q-1)}{3 M_I M_Q \left(A_I^2+A_Q^2\right)^{3/2}} \left[A_I F_1\left(\frac{3}{2};\frac{1}{2},m+1;\frac{5}{2};\frac{A_I^2}{A_I^2+A_Q^2},-\frac{2 (K+1)}{(A_I^2+A_Q^2) \gamma_0}\right)\right.\right.\\
        &\left.\left. + A_Q F_1\left(\frac{3}{2};\frac{1}{2},m+1;\frac{5}{2};\frac{A_Q^2}{A_I^2+A_Q^2},-\frac{2 (K+1)}{(A_I^2+A_Q^2) \gamma_0}\right)\right] \right. -\frac{\pi}{2 A_I^2}  \frac{M_I-1}{M_I}\\
        &\left. \times {}_2F_1\left(\frac{3}{2},m+1;2;-\frac{2 (K+1)}{A_I^2 \gamma_0}\right)  + \frac{\pi}{2 A_Q^2}  \frac{M_Q-1}{M_Q} {}_2F_1\left(\frac{3}{2},m+1;2;-\frac{2 (K+1)}{A_Q^2 \gamma_0}\right) \right\}
\end{aligned}$  & 
eq05 \\
\midrule
\begin{tabular}[c]{@{}l@{}}M-ary\\ SQAM\end{tabular} & 
$\!\begin{aligned}[t]
    P_s^{SQAM}(\gamma_0) &= \frac{2}{\pi} \frac{1+K}{\gamma_0} \frac{M-1}{9} \left(1-\frac{1}{\sqrt{M}}\right)  \sum_{m=0}^{\infty} \frac{(-1)^m}{m!}\left(\frac{K}{1+\Gamma^2}\right)^m {}_2F_1\left(-m,-m;1;\Gamma^2\right) \\
     & \times \left[3 \pi \, _2F_1\left(\frac{3}{2},m+1;2;-\frac{2 (K+1)(M-1)}{3 \gamma_0}\right) \right.\\
     &\left. - \sqrt{2} \left(1-\frac{1}{\sqrt{M}}\right) F_1\left(\frac{3}{2};\frac{1}{2},m+1;\frac{5}{2};\frac{1}{2},-\frac{(K+1) (M-1)}{ 3 \gamma_0}\right)\right]
\end{aligned}$ &                  
eq06 \\
\midrule
\begin{tabular}[c]{@{}l@{}}M-ary\\ ASK\end{tabular} & 
$\!\begin{aligned}[t]
    P_s^{ASK\phantom{B}}(\gamma_0) &= \frac{1+K}{\gamma_0} \sum_{m=0}^{\infty} \frac{(-1)^m}{m!} \left(\frac{K}{1+\Gamma^2}\right)^m {}_2F_1\left(-m,-m;1;\Gamma^2\right) \\ 
    & \times \frac{ (M^2-1)(M-1)}{6 M} {}_2F_1\left(\frac{3}{2},m+1;2;-\frac{(K+1)(M^2-1)}{3\gamma_0}\right)
\end{aligned}$ &                  
eq07 \\
\midrule
QPSK & 
$\!\begin{aligned}[t]
    P_s^{QPSK}(\gamma_0) &= \frac{1}{3\pi} \frac{1+K}{\gamma_0} \sum_{m=0}^{\infty} \frac{(-1)^m}{m!} \left(\frac{K}{1+\Gamma^2}\right)^m {}_2F_1\left(-m,-m;1;\Gamma^2\right) \\
    & \times \left[3 \pi \, _2F_1\left(\frac{3}{2},m+1;2;-\frac{2 (K+1)}{\gamma_0}\right) - \frac{1}{\sqrt{2}}  F_1\left(\frac{3}{2};\frac{1}{2},m+1;\frac{5}{2};\frac{1}{2},-\frac{K+1}{\gamma_0}\right)\right]
\end{aligned}$ &                  
eq08 \\
\midrule
BPSK & 
$\!\begin{aligned}
    P_s^{BPSK}(\gamma_0) = \frac{1+K}{4\gamma_0} \sum_{m=0}^{\infty} \frac{(-1)^m}{m!}\left(\frac{K}{1+\Gamma^2}\right)^m {}_2F_1\left(-m,-m;1;\Gamma^2\right) {}_2F_1\left(\frac{3}{2},m+1;2;-\frac{K+1}{\gamma_0}\right)
\end{aligned}$ &                  
eq09  \\
\midrule
\begin{tabular}[c]{@{}l@{}}M-ary\\ DPSK\end{tabular} & 
$\!\begin{aligned}[t]
    P_s^{DPSK}(\gamma_0)&=\sum\limits_{m=0}^{\infty} \frac{{}_2F_1\left(-m,-m;1,\Gamma^2\right)}{m!}  \left(\frac{-K}{1+\Gamma^2}\frac{\gamma_0 \sin^2{\frac{\pi}{M}}}{1+K+\gamma_0\sin^2{\frac{\pi}{M}}}\right)^m \\ 
    &\times \sum\limits_{p=0}^{\infty} \binom{m+p}{p}  \left(\frac{-(K+1)\cos\frac{\pi}{M}}{K+1+\gamma_0\sin^2{\frac{\pi}{M}}}\right)^p \left(\frac{m}{m+p}-\frac{\gamma_0 \sin^2{\frac{\pi}{M}}}{1+K+\gamma_0 \sin^2{\frac{\pi}{M}}}\right)\\
    &\times\left(\frac{\Gamma \left[\frac{p+1}{2}\right]}{2 \sqrt{\pi } \Gamma \left[\frac{p+2}{2}\right]}-\frac{\left(-\cos{\frac{Pi}{M}}\right)^{p+1}}{\pi  (p+1)}{}_2F_1\left(\frac{1}{2},\frac{p+1}{2};\frac{p+3}{2};\cos^2{\frac{\pi}{M}}\right)\right)
\end{aligned}$ &
eq10e \\
\bottomrule
\end{tabular}
\label{Tab:02}
\end{table*}

\begin{table*}
\caption{
Asymptotic ASEP expressions} 
\label{Tab:03}
\centering
\begin{tabular}{l|lL}
\toprule
\multicolumn{1}{c}{Modulation} & \multicolumn{2}{l}{ASEP} \\
\midrule
\begin{tabular}[c]{@{}l@{}}M-ary\\ RQAM\end{tabular}  & $\!\begin{aligned}[t]
     P_s^{RQAM}(\gamma_0) &\approx \frac{ 2 (M_I-1) (M_Q-1)}{ (A_I A_Q M_I M_Q)}\left[        \frac{A_Q}{A_I} \left(\tan ^{-1}(A_I,A_Q)+\frac{\pi }{2 (M_Q-1)}\right) ~~~~~~~~~~~~~~~~~~~~~~\phantom{1}\right.\\
     &-\frac{A_I}{A_Q}\left.\left(\tan ^{-1}(A_I,A_Q)-\frac{\pi  M_I}{2 (M_I-1)}\right)+1\right] \frac{K+1}{\pi  \gamma_0} e^{-K}  I_0\left(\frac{2 \Gamma K}{\Gamma^2+1}\right)
\end{aligned}$ & 
eq20 \\
\midrule
\begin{tabular}[c]{@{}l@{}}M-ary\\ SQAM\end{tabular} & 
$\!\begin{aligned}
     P_s^{SQAM}(&\gamma_0)\approx  \left(1+\frac{\pi(1+\sqrt{M})}{2(\sqrt{M}-1)}\right) \frac{2(\sqrt{M}-1)^2(M-1)}{3M}  \frac{K+1}{\pi  \gamma_0}  e^{-K} I_0\left(\frac{2 \Gamma K}{\Gamma^2+1}\right)
\end{aligned}$ &                  
eq21  \\
\midrule
\begin{tabular}[c]{@{}l@{}}M-ary\\ ASK\end{tabular} & $\!\begin{aligned}
     P_s^{ASK\phantom{B}}(&\gamma_0)\approx \frac{(K+1) (M-1)(M^2-1)}{6\gamma_0 M} e^{-K} I_0\left(\frac{2 \Gamma K}{\Gamma^2+1}\right)
\end{aligned}$ & 
eq22 \\
\midrule
QPSK & $\!\begin{aligned}
     P_s^{QPSK}(&\gamma_0)\approx \left(3+\frac{2}{\pi}\right)\frac{K+1}{4\gamma_0} e^{-K} I_0\left(\frac{2 \Gamma K}{\Gamma^2+1}\right)
\end{aligned}$ & 
eq23 \\
\midrule
BPSK & $\!\begin{aligned}
     P_s^{BPSK}(&\gamma_0)\approx \frac{K+1}{4\gamma_0} e^{-K} I_0\left(\frac{2 \Gamma K}{\Gamma^2+1}\right)
\end{aligned}$ & 
eq24 \\
\midrule
\begin{tabular}[c]{@{}l@{}}M-ary\\ DPSK\end{tabular} & 
$\!\begin{aligned}
     P_s^{DPSK}(&\gamma_0)\approx \frac{1}{M} \frac{(K+1)(M-1)}{K+1+\gamma_0\sin^2{\frac{\pi}{M}}} e^{\frac{-K\gamma_0\sin^2{\frac{\pi}{M}}}{K+1+\gamma_0\sin^2{\frac{\pi}{M}}}} I_0\left(\frac{2\Gamma}{1+\Gamma^2}\frac{K\gamma_0\sin^2{\frac{\pi}{M}}}{K+1+\gamma_0\sin^2{\frac{\pi}{M}}}\right) 
\end{aligned}$ & 
eq25 \\
\bottomrule
\end{tabular}
\end{table*}

\subsection{Asymptotic average symbol error probability expressions}
To effectively analyze the influence of TWDP fading on the achievable error rate performance of considered modulation techniques, the corresponding concise asymptotic closed-form ASEP expressions in the high SNR region are also derived, enabling us to relax the computational complexity which occurs for large values of $K$. Despite its importance, to the best of authors' knowledge, the asymptotic analysis is available only for BPSK~\cite{Kim17, rad} and DBPSK~\cite{Rao15} modulations with no diversity, while for all other modulation  techniques these expression are missing.

\subsubsection{M-ary RQAM and its derivatives}
So, considering that \mbox{$F_1\left(a; b_1,b_2;c;z_1,z\right)\sim {}_2 F_1\left(a,b_1;c;z_1\right)$} and \mbox{${}_2 F_1\left(a,b;c;z\right)\sim 1$} when \mbox{$z\xrightarrow{} 0$}, equation (\ref{eq05}) for large values of $\gamma_0$ can be expressed as (\ref{eq20}),
which represents the simple, closed-form asymptotic RQAM ASEP expression. By inserting the appropriate values of $M_I$, $M_Q$, $A_I$ and $A_Q$ in (\ref{eq20}) given by Table~\ref{table:RQAM}, the closed-form asymptotic ASEP expressions are also obtained for M-ary SQAM, M-ary ASK, QPSK and BPSK, and are given by (\ref{eq21}) - (\ref{eq24}), respectively (see Table~\ref{Tab:03}).

\subsubsection{M-ary DPSK}
In order to derive the asymptotic M-ary DPSK expression, one can show that in (\ref{eq10e}) $\sum_{p=0}^{\infty}(\cdot) \sim \left(1-\frac{1}{M}\right) \frac{1+K}{1+K+\gamma_0 \sin^2{\frac{\pi}{M}}}$ when $\gamma_0 \to \infty$, which can be, according to 
~\cite[eq. (23)]{rad}, reduced to a final form of a M-ary DPSK asymptotic ASEP given by (\ref{eq25}) in Table~\ref{Tab:03}.



\section{Numerical results and discussion}
\label{sec:IV}
In this section, we compared analytical results of the exact and asymptotic analysis conducted in Section~\ref{sec:III} with the results obtained using Monte-Carlo simulation.

\subsection{Exact analysis}
\begin{figure}[!h]
	\captionsetup[subfigure]{aboveskip=-3pt,belowskip=-4pt}
	\centering
	\begin{subfigure}{0.5\textwidth}
		\centering
		\includegraphics[width=1\textwidth]{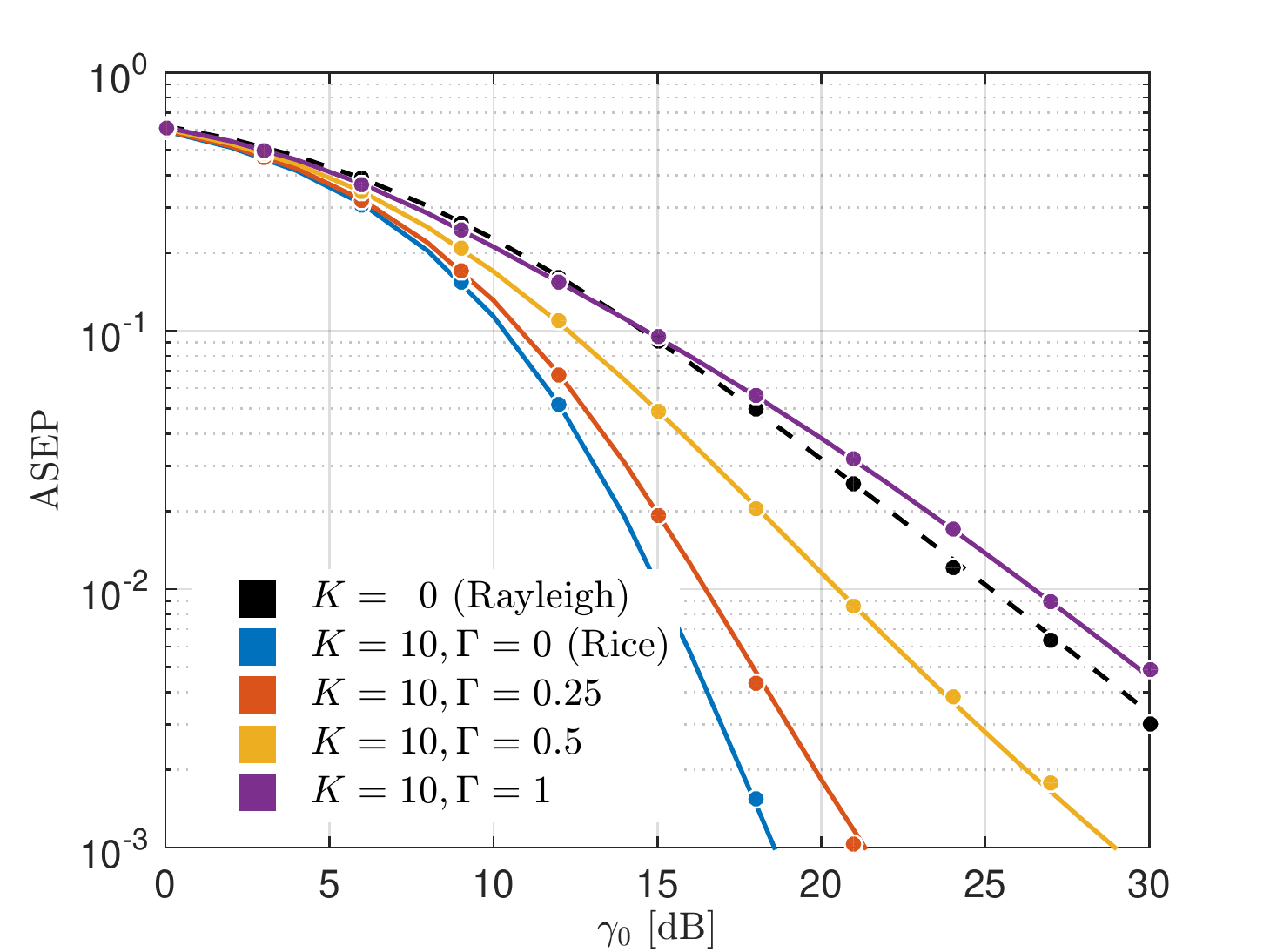}
		\caption{}
		\label{Fig1a}
	\end{subfigure}
	\begin{subfigure}{0.5\textwidth}
		\centering
		\includegraphics[width=1\textwidth]{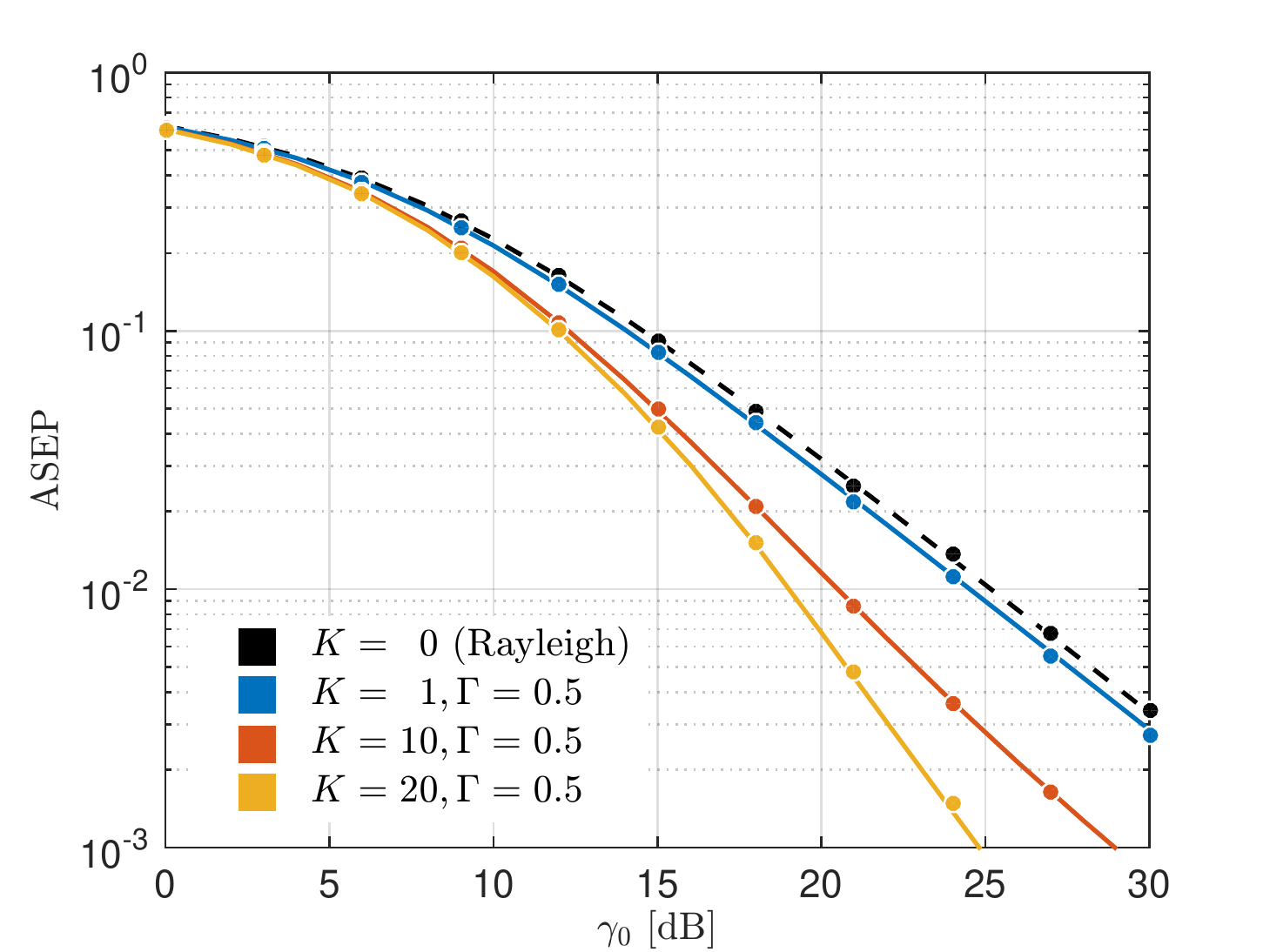}
		\caption{}
		\label{Fig1b}
	\end{subfigure}
	\begin{subfigure}{0.5\textwidth}
		\centering
		\includegraphics[width=1\textwidth]{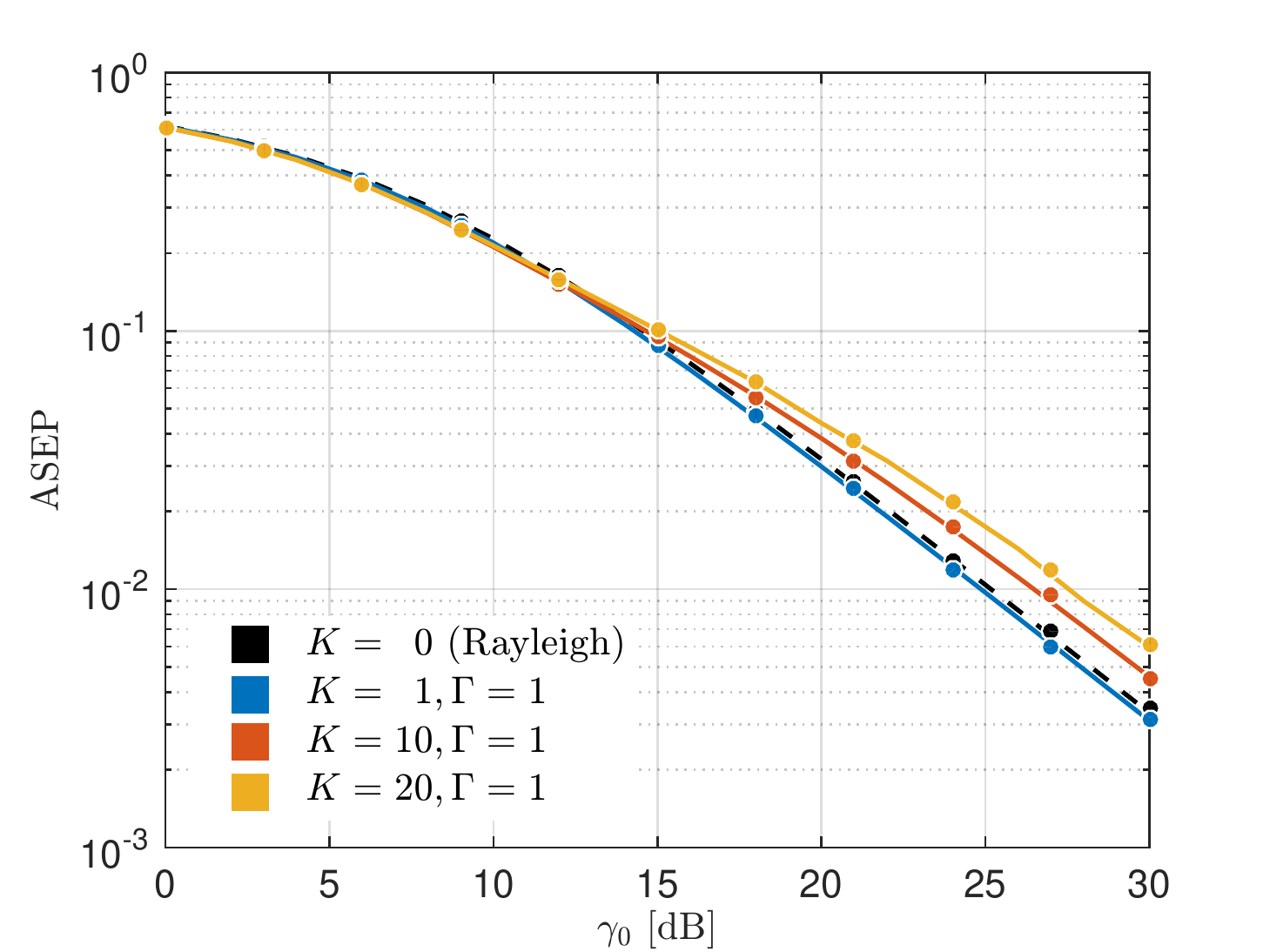}
		\caption{}
		\label{Fig1c}
	\end{subfigure}
	\caption{Analytical (solid line) and simulated (dots) ASEP vs. average symbol SNR for 4x2 RQAM modulation}
	\label{Fig1}
\end{figure}

\begin{figure}
	\captionsetup[subfigure]{labelformat=empty,aboveskip=-1pt,belowskip=-1pt}
	\centering
	\includegraphics[width=0.5\textwidth]{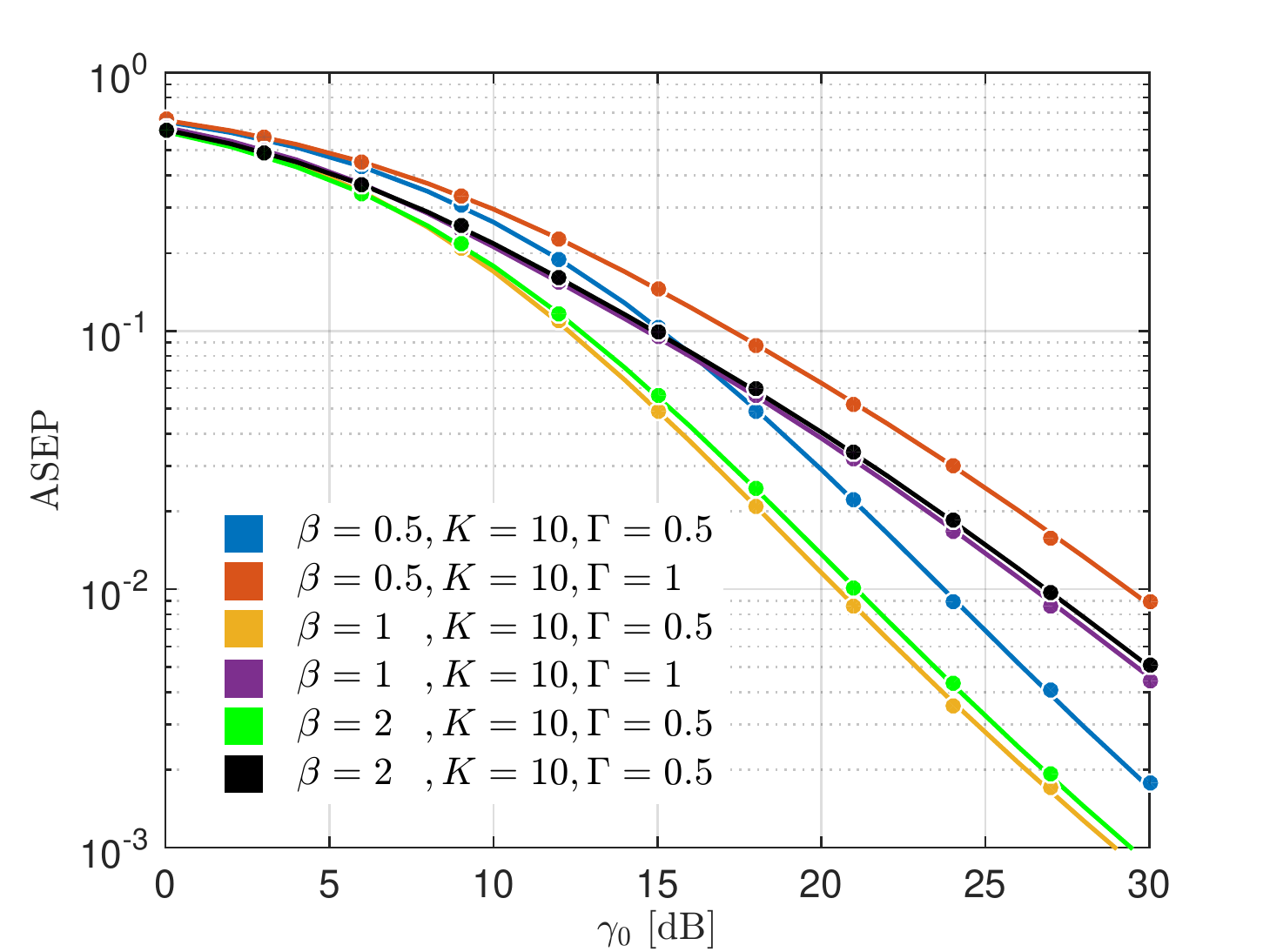}
	\caption{Analytical (solid line) and simulated (dots) ASEP vs. average symbol SNR for 4x2 RQAM modulation, for different combinations of $(K, \Gamma)$ and different values of $\beta$}
	\label{Fig3}
\end{figure}

\begin{figure*}
	\begin{subfigure}{.5\textwidth}
		\centering
		\includegraphics[width=\textwidth]{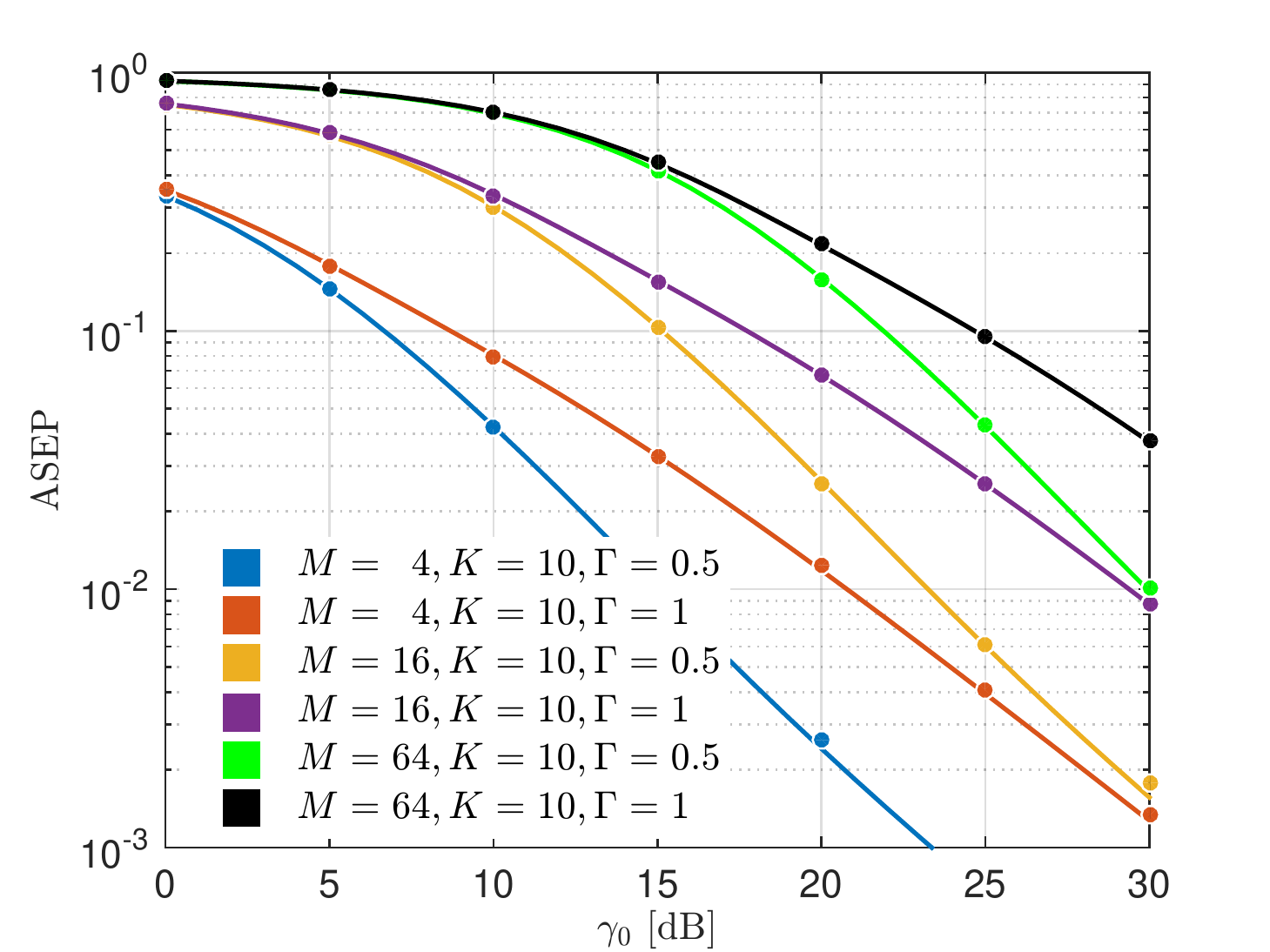}
		\caption{}
		\label{Fig2b}
	\end{subfigure}
	\begin{subfigure}{.5\textwidth}
		\centering		\includegraphics[width=\textwidth]{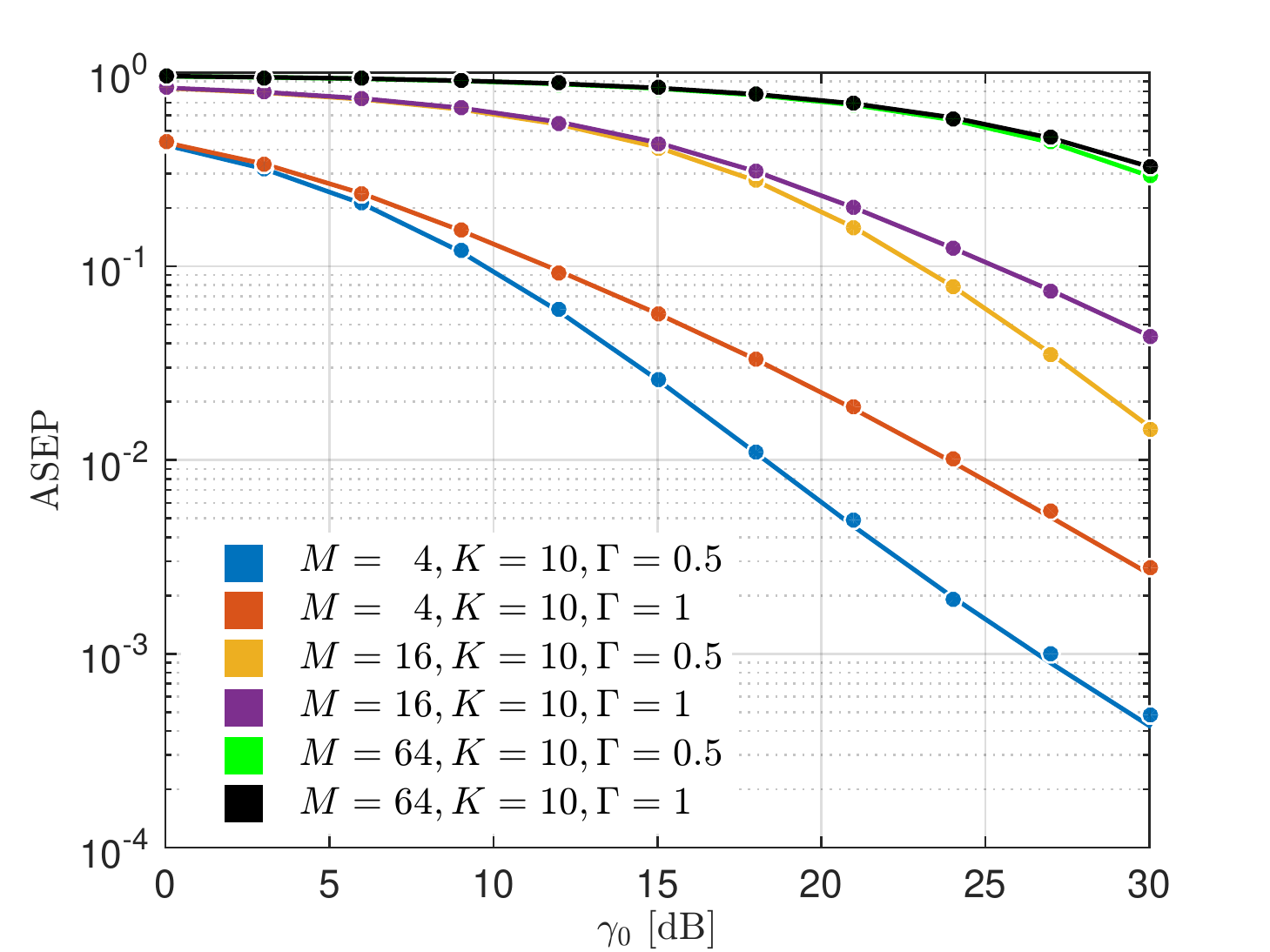}
		\caption{}
		\label{Fig2a}
	\end{subfigure}
	\newline
		\begin{subfigure}{.5\textwidth}
		\centering
		\includegraphics[width=\textwidth]{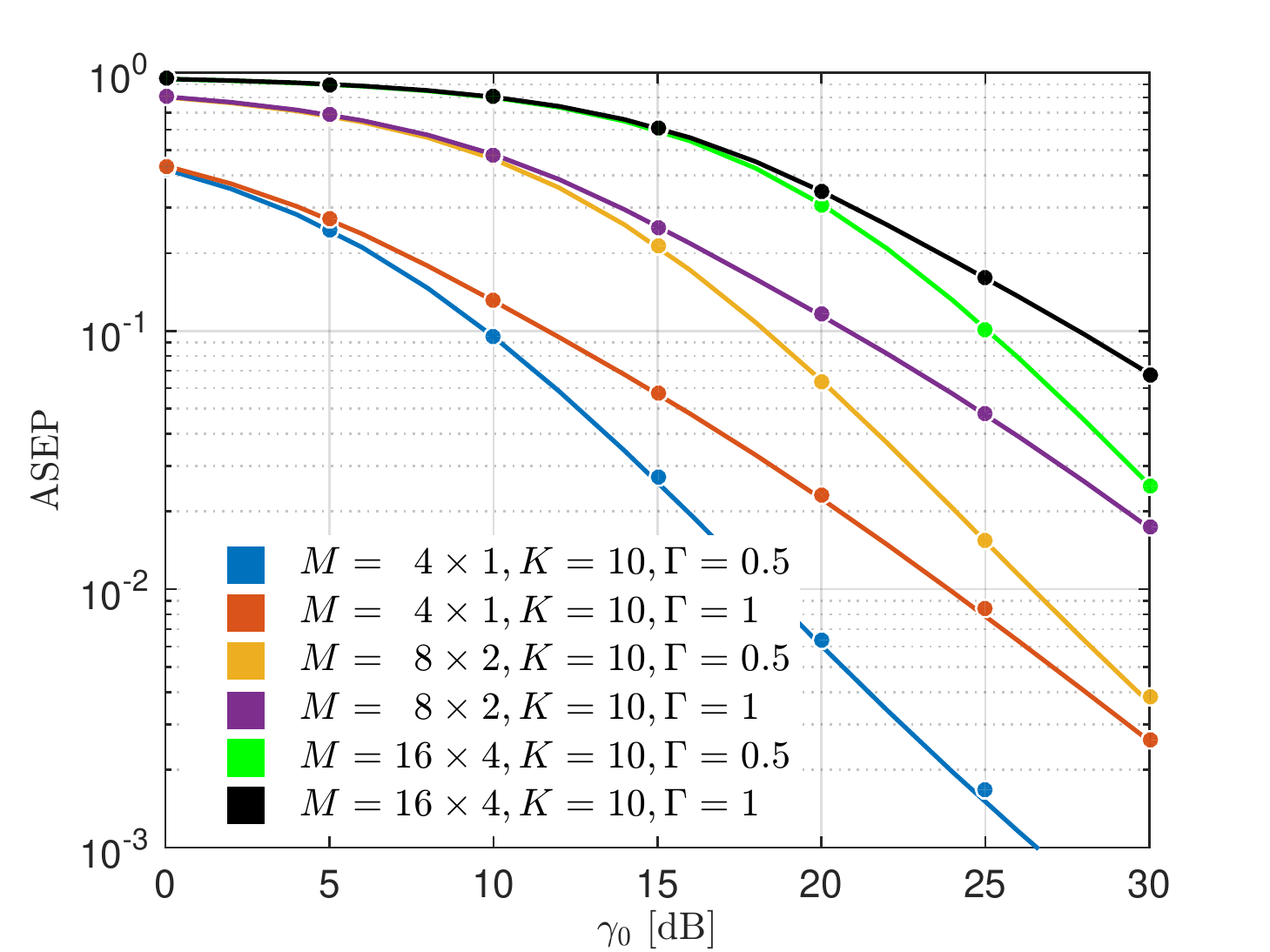}
		\caption{}
		\label{Fig2c}
	\end{subfigure}
	\begin{subfigure}{.5\textwidth}
		\centering
		\centering		\includegraphics[width=\textwidth]{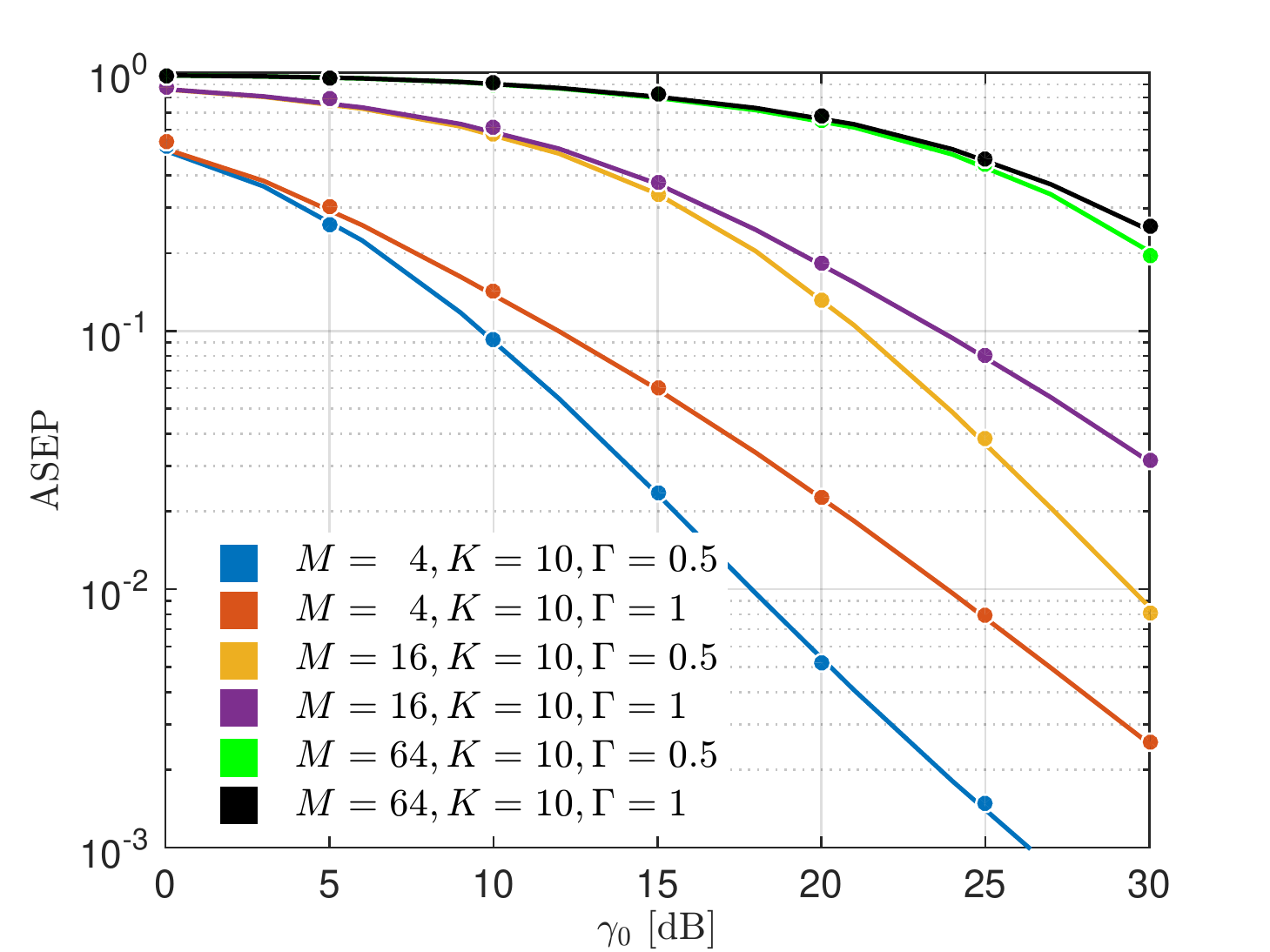}
		\caption{}
		\label{Fig2d}
	\end{subfigure}
	\caption{Analytical (solid line) and simulated (dots) ASEP vs. average symbol SNR for (a) M-ary SQAM (b) M-ary ASK, (c) M-ary RQAM  and (d) M-ary DPSK modulation}
	\label{Fig2}
\end{figure*}

In order to verify derived exact ASEP expressions and to investigate the impact of different modulation orders, schemes and fading parameters on error performance in TWDP channels, (\ref{eq05}) - (\ref{eq07}) and (\ref{eq10e}) versus average SNR 
are illustrated in Fig.~\ref{Fig1} - Fig.~\ref{Fig2}, together with the results obtained by the corresponding Monte-Carlo simulations generated using $10^{5}$ samples. 
An excellent match between analytical and simulated results, for all considered modulation orders  $M$ and TWDP parameter tuples $(K, \Gamma)$, can be observed from all the figures. That validates derived theoretical expressions summarized in Table~\ref{Tab:01} and enable us to make conclusions about the behaviour of differently modulated signals in different TWDP fading channels.

 Accordingly, the impact of different fading conditions (expressed by different values of a tuple $(K, \Gamma)$) on ASEP are illustrated in Fig.~\ref{Fig1}, for 4x2 RQAM modulation (with $\beta = 1$). From Fig.~\ref{Fig1a} can be observed that for a fixed value of $K$, performances degrade with the increasement of $\Gamma$, ranging from those obtained in Rician channel with the same $K$, to those obtained in hyper-Rayleigh channels. From Fig.~\ref{Fig1b} can also be observed that 
for a small and moderate values of $\Gamma$ (e.g. $\Gamma = 0.5$), error probability reduces as $K$ grows. Conversely, Fig.~\ref{Fig1c} indicates that for a large values of $\Gamma$ ($\Gamma \to 1$), error probability increases with the increasement of $K$,
finally becoming worse that the one obtained for Rayleigh channel.

Fig.~\ref{Fig1a} also reveals benefits of  $\Gamma$-based parameterization in respect to $\Delta$-based. Namely, by comparing ASEP curves given in Fig.~\ref{Fig1a} with those given in~\cite[Fig. 4]{Rao14} (for BPSK), it can be observed that for the same absolute values of parameters $\Delta$ and $\Gamma$ ($\Delta \in \{0, 0.25, 0.5, 1\}$ and $\Gamma \in \{0, 0.25, 0.5, 1\}$), $\Delta$-parameterized ASEP curves are almost indistinguishable when $\Delta$ changes its values from 0 to 0.5. On the contrary, $\Gamma$-parameterized
ASEP curves differ notably, enabling us to clearly observe the impact of the increment of the ratio between specular components on error performances in TWDP channels. 

The impact of different ratios between $d_I$ and $d_Q$ within RQAM, i.e $\beta \in [0.5, 1, 2]$, on ASEP, is illustrated in Fig.~\ref{Fig3}, which presents ASEPs of 4x2 RQAM for ($K = 10$ and $\Gamma = 0.5$) and ($K = 10$ and $\Gamma = 1$). The figure shows that for a considered channel conditions, minimum ASEP is achieved for $\beta = 1$, i.e. when in-phase and quadrature distances are the same. 

So, considering the results provided in Fig.~\ref{Fig3}, Fig.~\ref{Fig2} enables us to compare the best case RQAM ASEP obtained for $\beta = 1$, with the results obtained for M-ary SQAM, M-ary ASK and M-ary DPSK for the same values of $M \in \{4, 16, 64\}$ and for different values of a tuple $(K, \Gamma)$ i.e. ($K = 10$ and $\Gamma = 0.5$) and ($K = 10$ and $\Gamma = 1$). Accordingly, Fig.~\ref{Fig2b} - Fig.~\ref{Fig2d} revel superiority of SQAM in respect to generalized RQAM, ASK and DPSK, since it provides minimum ASEP for the same value of SNR. Since the aforesaid can be observed for all considered fading conditions and modulation orders, one can conclude that SQAM modulation has the highest energy efficiency among the others treated, which, together with its high bandwidth efficiency and its implementation simplicity, justifies the widespread applicability of SQAM modulation in highly-demanding 5G networks. 

\subsection{Asymptotic analysis}
\begin{figure}[!h]
	\captionsetup[subfigure]{aboveskip=-4pt,belowskip=-4pt}
	\centering
	\begin{subfigure}{0.5\textwidth}
		\centering		\includegraphics[width=1\textwidth]{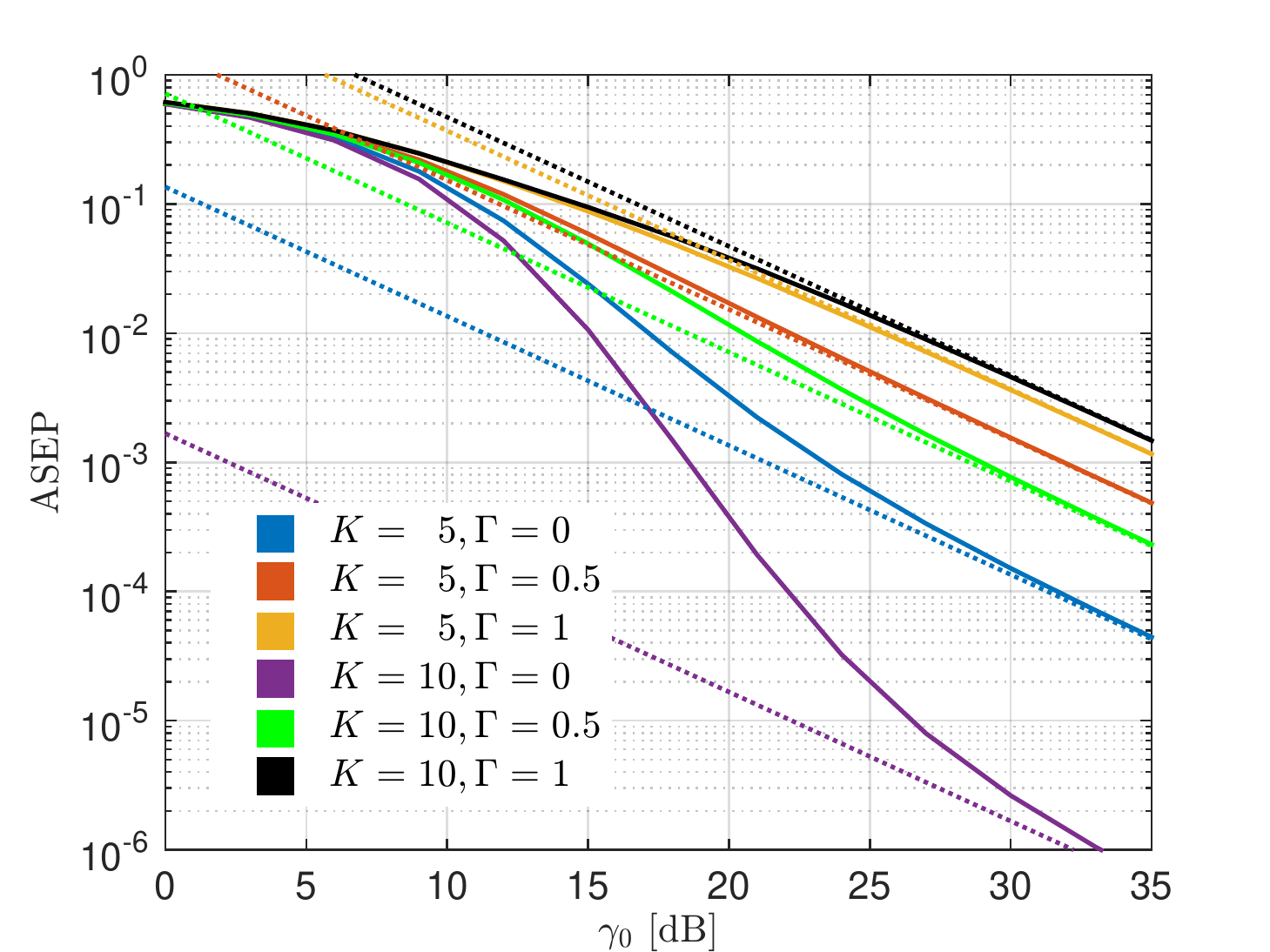}
		\caption{}
		\label{Fig5a}
	\end{subfigure}
	\begin{subfigure}{0.5\textwidth}
		\centering
		\includegraphics[width=1\textwidth]{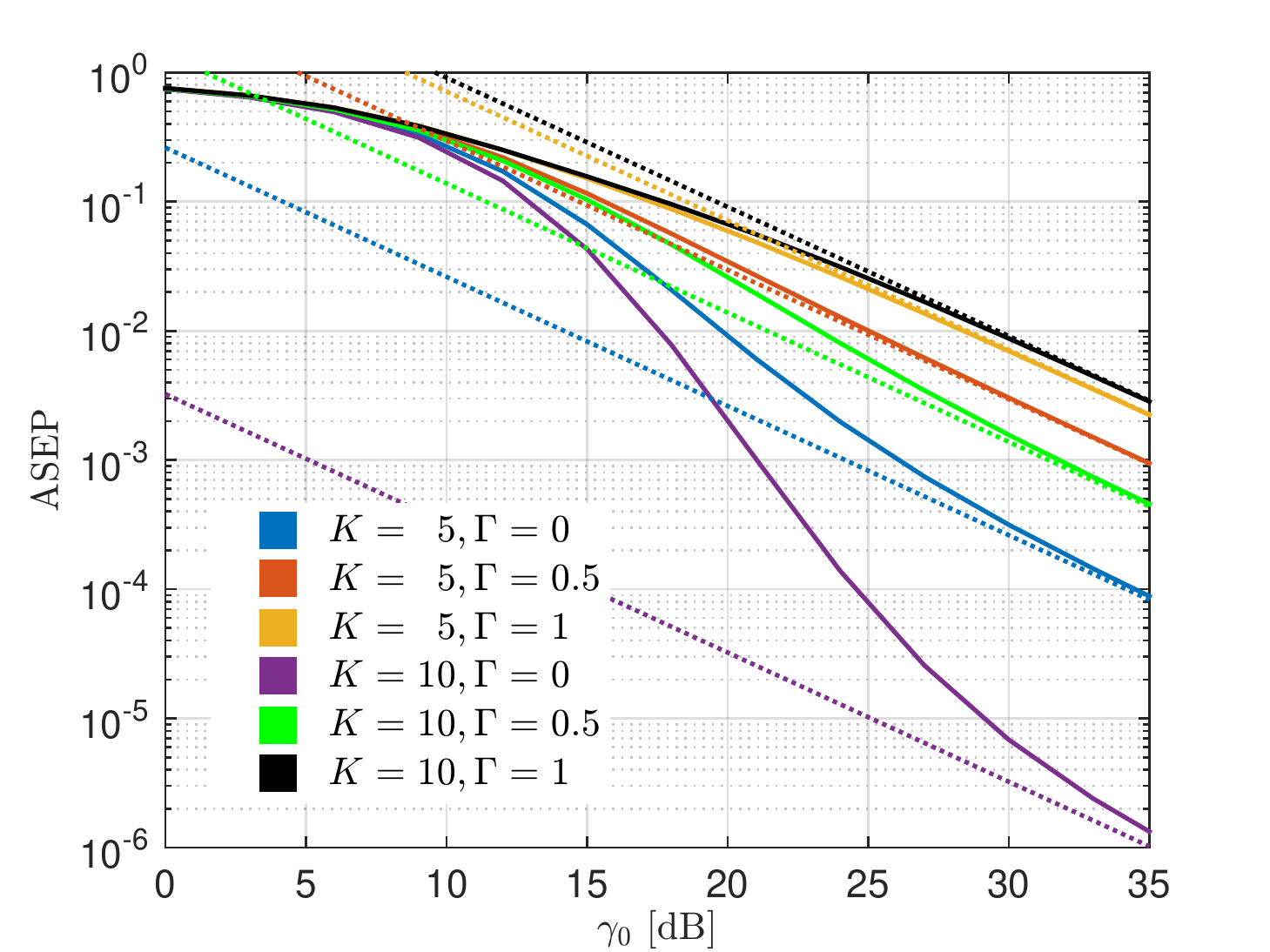}
		\caption{}
		\label{Fig5b}
	\end{subfigure}
	\begin{subfigure}{0.5\textwidth}
		\centering
		\includegraphics[width=1\textwidth]{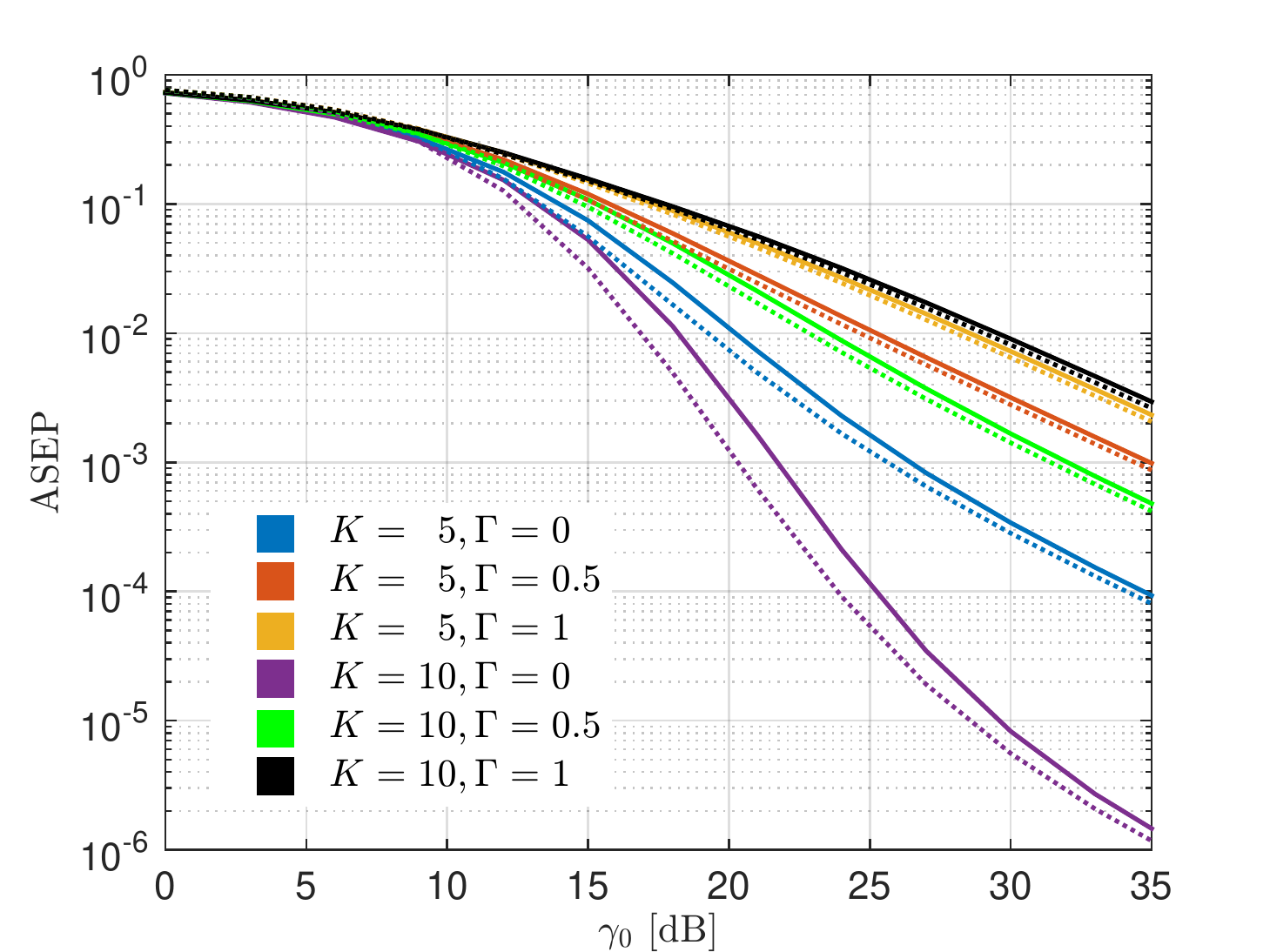}
		\caption{}
		\label{Fig5c}
	\end{subfigure}
		\caption{Asymptotic (dotted line) and the exact (solid line) ASEP vs. average symbol SNR for (a) 4x2-RQAM, (b) 16-SQAM and (c) 8DPSK modulations}
	\label{Fig5}
\end{figure}

At the end of this Section, we also provide numerical results to validate performed asymptotic error performance analysis, with its results summarized in Table~\ref{Tab:03}. Accordingly, Fig.~\ref{Fig5} illustrates the existing analytical ASEP expressions (i.e. (\ref{eq05}), (\ref{eq06}) and (\ref{eq10e})) obtained for 4x2-RQAM, 16-SQAM and 8DPSK and $K \in [5, 10]$ and $\Gamma \in [0, 0.5, 1]$.
Additionally, analytical results
and those obtained from asymptotic closed-form expressions given by (\ref{eq20}),  (\ref{eq21}) and (\ref{eq25}) are also presented for considered modulation orders and values of a tuple $(K, \Gamma)$,
enabling us to compare the exact and the asymptotic expressions in different propagation scenarios, ranging from better-then-Rice to worse-than-Rayleigh fading conditions.

From Fig.~\ref{Fig5} can be observed that the derived asymptotic results show great agreement with analytical ASEP curves in the high SNR region (i.e. for SNR>30~dB). It also can be noticed that the results are the worst in terms of SNR lower bound for channel conditions which can be described also by using Rician distribution (i.e. for $\Gamma = 0$), while for conditions which can only be described by TWDP distribution, asymptotic and the exact ASEP expressions remarkably agree even for considerably smaller values of SNR (typically greater then SNR>20~dB).
Accordingly, derived asymptotic ASEP expressions summarized in Table~\ref{Tab:03} can be effectively utilized to provide the explicit insights into the achievable error rate performance of RQAM, SQAM and DPSK modulated signal over the TWDP fading
channels for high SNR, for varieties of modulation orders and fading conditions.

\section{Conclusion}
Despite the prevalence of TWDP model in description of a small-scale fading effects in mmWave band and cavity sensor networks, the exact ASEP expressions 
are to date provided only for M-ary PSK and M-ary FSK. 
Accordingly, in this paper, the exact and high-SNR asymptotic ASEP expression are derived for 
M-ary RQAM and M-ary SQAM with coherent detection, as well as for differential detected M-ary PKS modulation, thus closing the gap by providing the exact ASEP expressions for the most popular modulations with the single-antenna reception. Analytical results are shown to be in perfect accordance with simulated results, thus providing the appropriate tool for accurate analytical evaluation of error performances in TWDP fading channels.    

\label{sec:V}

\section*{Acknowledgment}
The authors would like to thank Prof. Ivo Kostić for many valuable discussions and advice.

\bibliographystyle{IEEEtran}
\bibliography{Literatura}



\end{document}